\documentclass[11pt,tightenlines,eqsecnum,floats,aps,amsmath,amssymb,nofootinbib,prd,shownopacs,floatfix]{revtex4}

\usepackage{epstopdf}
\usepackage{latexsym}
\usepackage{graphicx}
\usepackage{amsmath}
\usepackage{amssymb}
\usepackage{mathrsfs}
\usepackage{xparse}
\usepackage{mathtools}
\usepackage{wasysym}
\usepackage{fancyhdr}
\usepackage[utf8]{inputenc}
\usepackage{mathrsfs}
\usepackage{soul}
\usepackage{float}
\usepackage[center]{subfigure}
\usepackage{color}
\usepackage{xparse}
\usepackage{appendix}

\begin{document}
\title{Quantum gravitational non-singular tunneling wavefunction proposal}

\author{Meysam Motaharfar}
\email{mmotah4@lsu.edu}

\author{Parampreet Singh}
\email{psingh@lsu.edu}
\affiliation{Department of Physics and Astronomy, Louisiana State University, Baton Rouge, LA 70803, USA}


\begin{abstract}
It was recently shown that tunneling wavefunction proposal is consistent with loop quantum geometry corrections including both holonomy and inverse scale factor corrections in the gravitational part of a spatially closed isotropic model with a positive cosmological constant. However, in presence of an inflationary potential the initial singularity is kinetic dominated and the effective minisuperspace potential again diverges at zero scale factor.  Since the wavefunction in loop quantum cosmology cannot increase towards the zero scale factor, the tunneling wavefunction seems incompatible. We show that consistently including inverse scale factor modifications in  scalar field Hamiltonian changes the effective potential into a barrier potential allowing the tunneling proposal. We also discuss a potential quantum instability of the cyclic universe resulting from tunneling.
\end{abstract}

\maketitle

\section{Introduction}

\begin{figure}
    \centering
    \includegraphics[scale = 0.7]{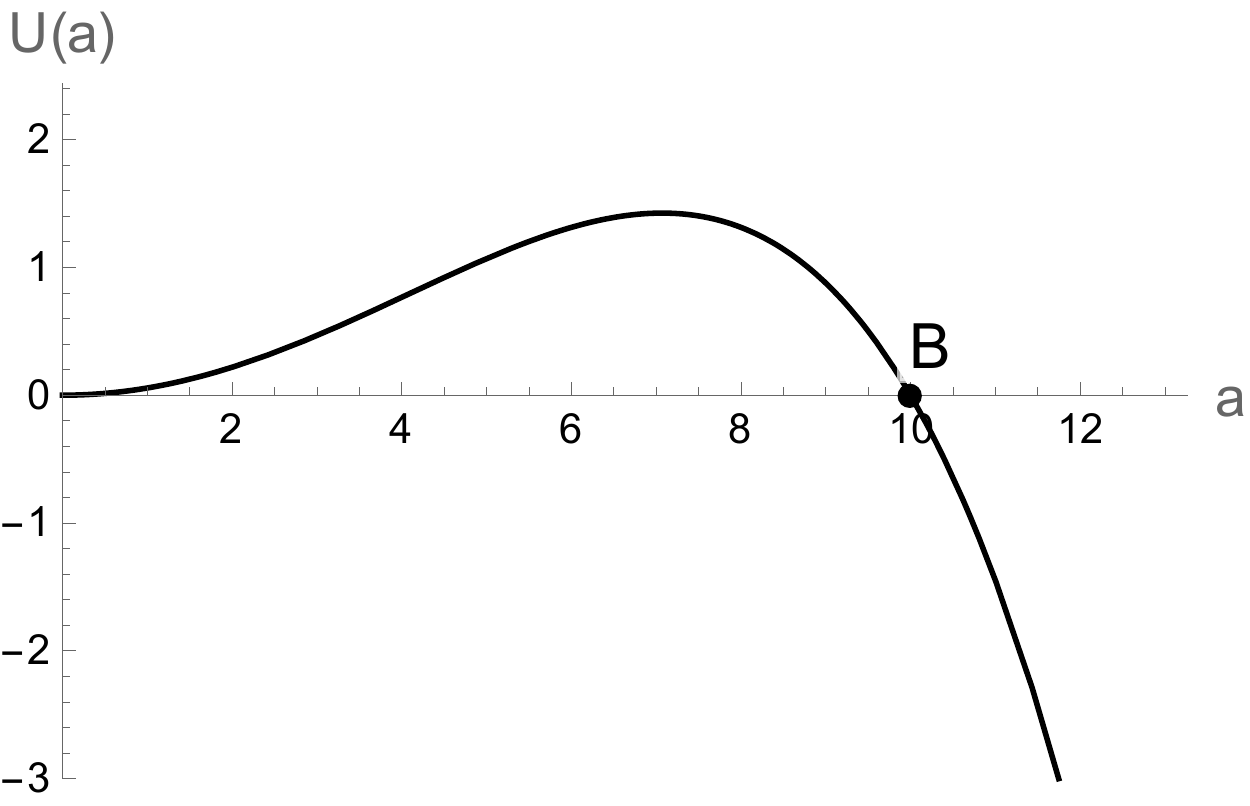}
    \caption{Effective minisuperspace potential for Wheeler-DeWitt quantum cosmology. We set $G=1$ and $\Lambda = 0.03$.}
    \label{V}
\end{figure}

Did the universe have a beginning? The answer to this question is in affirmative in the classical theory due to singularity theorems proved by Penrose, Hawking and Geroch in which they demonstrated that considering reasonable energy conditions, the universe must have begun from a big bang singularity in the past \cite{Geroch:1968ut, Hawking:1970zqf, Hawking-Ellis} (see also Ref. \cite{Borde:2001nh} for recent version of singularity theorem in inflationary cosmology). In classical general relativity (GR), singularities are the boundary of spacetime where all physical laws break down indicating the need for new physics. Since the entire cosmos should be treated as a closed quantum system, the boundary conditions must be supplied as part of dynamical laws. Several proposals have been put forward  to describe the boundary conditions of the wavefunction of the universe leading to a self-contained universe among which tunneling wavefunction proposal \cite{Vilenkin:1982de, Vilenkin:1984wp} and no-boundary wavefunction proposal \cite{Hartle:1983ai, Hawking:1983hj} are the most successful ones. Although these proposals were formulated differently, they both can be described using the Wheeler-DeWitt quantum cosmology as it was first studied in Ref. \cite{Vilenkin:1987kf}. Considering closed isotropic universe with cosmological constant $\Lambda$, one can write the Wheeler-DeWitt equation as follows 
\begin{align}\label{Wheeler-DeWitt}
    \left [a^{-n} \frac{d}{da} a^{n} \frac{d}{da} - U(a)\right] \Psi(a) = 0,
\end{align}
where $a$ is the scale factor and the parameter $n$ represents factor ordering ambiguity, and $U(a)$ is the effective minisuperspace potential
\begin{align}\label{potential}
U(a) = \frac{36}{\kappa^2} a^2 \left(1-  \frac{8\pi G}{3} \rho(a) a^2\right),
\end{align}
where $\kappa = 8\pi G$ and $\rho = \Lambda/ (8\pi G)$. Looking at Eq. (\ref{Wheeler-DeWitt}), one can see the similarity of Wheeler-DeWitt equation with Schrodinger equation with zero energy eigenvalue. Moreover, from Fig. \ref{V}, one can see that the effective minisuperspace potential has two regimes; classically forbidden regime (so-called Eucleadian regime) and classically allowed regime. In fact, the barrier shape of effective minisuperspace potential manifests the analogy of creation of the universe out of nothing via quantum tunneling phenomena. Classically, the universe contracts from large size, bounces and expands. While quantum mechanically, the universe can start at zero scale factor with zero energy, i.e. nothing (``nothing" here means no space, time, and matter), and then tunnel through the barrier to classical expanding universe. Since the potential has a barrier shape, the wavefunction is a superposition of growing and decreasing wave modes under the barrier while it a superposition of oscillatory wave modes outside the barrier. In analogy with quantum tunneling, tunneling wavefunction describes the boundary conditions while the wavefunction has increasing wave mode towards the big bang and has just out going wave mode outside the barrier like a particle escape the radioactive nucleus. Using the WKB approximation, the wavefunction is given by \cite{Vilenkin:1987kf}    
\[   
\Psi_{V}(a) = 
     \begin{cases}
       e^{\int^{B}_{a} \sqrt{|U(a^{\prime})|} da^{\prime}} &  \ \ \ a < B \\
       e^{-i\int^{a}_{B} \sqrt{|U(a^{\prime})|} da^{\prime} + i\frac{\pi}{4}} &  \ \ \ a\ge B
\end{cases}
\]
\\
where $B = \sqrt{{3}/{\Lambda}}$ is the scale factor at which the universe bounces classically. However, the no-boundary proposal describes the boundary conditions of universe in such a way that the wavefunction has decreasing mode towards the big bang under barrier and it is in superposition of in going (contracting universe) and outgoing (expanding universe) wave modes outside the barrier. Similarly, the wavefunction is given by \cite{Vilenkin:1987kf} 
\[   
\Psi_{HH}(a) = 
     \begin{cases}
       e^{-\int^{B}_{a} \sqrt{|U(a^{\prime})|} da^{\prime}} &  \ \ \ a < B\\
       \cos(\int^{a}_{B} \sqrt{|U(a^{\prime})|} da^{\prime} - \frac{\pi}{4}) &  \ \ \ a\ge B
\end{cases}
\]
\\
where the wavefunction is real which is a property of no-boundary wavefunction. Given the wavefunction, the nucleation probability for which the universe tunnels from nothing into classical expanding universe reads as
\begin{align}\label{Psi}
    P_{V, HH}  \sim e^{\pm \frac{c}{\Lambda}}
\end{align}
where $c$ is a positive constant and positive (negative) sign stands for no-boundary (tunneling) proposal. From Eq. (\ref{Psi}), one can see that the nucleation probability peaks at smaller value of cosmological constant in case of no-boundary proposal meaning that the universe favors tunneling to large expanding universe while the opposite is true for tunneling boundary proposal. On the other hand, this means that no-boundary proposal has the largest probability for inflation to happen at the minimum of potential while tunneling wave function has the largest probability for inflation to occur at the top of potential which is theoretically and observationally favored. However, it was discussed that no-boundary proposal also predicts large amount of inflation after multiplying nucleation probability by volume weighting \cite{Hartle:2007gi}.

Although these boundary proposals were successful in describing the boundary conditions of the universe, they are based on semi-classical physics. However, one must consider quantum gravity effects when the universe reaches the Planck regime. In fact, it is a reasonable question to ask how the effective minisuperspace potential as well as boundary proposals get modified in the presence of quantum gravity effects. This issue was investigated in Ref. \cite{Motaharfar:2022pjp} for tunneling wavefunction proposal for a spatially closed universe in Loop Quantum Cosmology (LQC) with a positive cosmological constant.\footnote{See Refs. \cite{Brahma:2018elv, Brahma:2018kkr} for discussion about no-boundary proposal in LQC.} This analysis assumed the validity of effective spacetime description in LQC at all scales resulting in a modified Friedmann dynamics where quantum geometry effects originate from holonomy modifications and inverse scale factor modifications. In the non-compact spatially-flat models the latter do not contribute, but they can be included in the spatially-compact and spatially-curved models. In fact, for spatially-curved anisotropic models inverse scale factor effects play an important role on obtaining bounds on anisotropic shear \cite{Gupt:2011jh}. It was found that taking into account just holonomy correction, the effective minisuperspace potential is lifted up at zero scale factor and the tunneling proposal becomes incompatible. However, at small scale one needs to consider the inverse scale factor correction together with holonomy correction due to which the effective minisuperspace potential again recovers its barrier shape and the universe can also pick up the tunneling boundary conditions. Moreover, it was found that the universe can tunnel from nothing into either a classical expanding universe or quantum cyclic universe depending on how large the cosmological constant is.

Above results are valid only for the case of pure cosmological constant but for inflation one obtains kinetic dominated phase as the singularity is approached \cite{foster}. The energy density behaves as that of massless scalar field as as $a^{-6}$, and in absence of any quantum, geometric modifications to energy density the effective minisuperspace potential will diverge at zero scale factor seemingly making the tunneling wavefunction proposal inconsistent with loop quantum geometry effects. On the other hand, in case of cyclic universe, the universe can also tunnel back from bounce into zero scale factor indicating quantum instability of cyclic universe of this type in LQC. To understand this issue let us note this instability in Wheeler-DeWitt case. The effective minisuperspace potential Eq. (\ref{potential})  is approximated by $U(a) \sim \rho(a) a^4$ in small scale factor regime. This means that any cyclic universe constructed using cosmological constant and perfect fluid with $\omega<1/3$ is not stable against tunneling to zero scale factor in Wheeler-DeWitt quantum cosmology \cite{Graham:2011nb, Mithani:2011en, Mithani:2012ii}. However, it was discussed in Ref. \cite{Graham:2014pca} that the effective minisuperspace potential for such cyclic universe may get modified at small scale with Casimir energy with $\omega  = 1/3$ lifting up the effective minisuperspace potential at zero scale factor stabilizing the cyclic universe against non-perturbative decay towards vanishing size. Therefore, those cyclic universes which are built from matter with $\omega\ge 1/3$ are stable against tunnel to nothing in Wheeler-DeWitt quantum cosmology. However, the effective Friedmann equation in LQC gets modified with quadratic energy density at high energy limit, so the effective minisuperspace potential is approximated by $U_{eff}(a) \sim \rho^2 a^4$ \cite{Motaharfar:2022pjp}. Therefore, it seems that cyclic universe with $\omega\ge -1/3$ may be stable against quantum decay to vanishing size in LQC. However, it was discussed in Ref. \cite{Mithani:2014jva} that even one considers massless scalar field, i.e., $\omega=1$, there is finite probability to tunnel back to zero scale factor for emergent/cyclic universe build in the context of LQC. 

These results at first seem to be in contradiction with what was inferred from effective minisuperspace potential found in Ref. \cite{Motaharfar:2022pjp} if one includes massless scalar field or any perfect fluid with $\omega\ge -1/3$. However, we show that this inconsistency stems from ignoring in the small-scale factor regime  effect of inverse scale factor correction for matter content. Therefore, the purpose of this manuscript is twofold. First, considering the cosmological constant plus massless scalar field to mimic the dynamics of inflationary cosmology to investigate the possibility of creation of the universe out of nothing into inflationary universe via tunneling wavefunction proposal considering the inverse scale factor correction for massless scalar field. Second, to study the quantum stability of cyclic universes, constructed using cosmological constant and perfect fluid, in the context of LQC. An important caveat in including inverse scale factor modifications is that in the regime when inverse scale factor effects can play any role the quantum fluctuations can be large and effective description might become a suspect. But surprisingly, the effect of large quantum fluctuations is to lower the density at which the bounce occurs in LQC \cite{Corichi:2011rt, Diener:2014hba}. In fact, for such states the modified Friedmann dynamics is still valid with the only change in lowering the bounce density \cite{Ashtekar:2015iza}. Though the above results are obtained for spatially-flat model, they are quite relevant for the model under consideration because at small scale factors the spatial curvature does not dominate in comaprison to to the energy density.
In the next section, we discuss the effective dynamics of spatially closed LQC. Then we derive the effective minisuperspace potential including both holonomy and inverse scale factor corrections in section \ref{III} and discuss how adding massless scalar field may change the effective minisuperspace potential at small scale regime. Finally, we give a summary of the results and conclusion.  

\section{k=1 loop quantum cosmology: effective dynamics}

The canonical quantization in LQG is based on using Ashtekar-Barbero variables due to which one can express the field strength of the connection in the Hamiltonian constraint in terms of the holonomies of the connection which are computed over a loop with a minimum area determined by the quantum geometry. Applying LQG techniques to symmetry reduced isotropic universe, one obtains a quantum Hamiltonian constraint which turns out to be a difference equation which results in singularity resolution  
\cite{Ashtekar:2006wn, Ashtekar:2006rx, Ashtekar:2006es, Ashtekar:2006uz, Ashtekar:2007em, Ashtekar:2011ni}. 
Interestingly the quantum dynamics in LQC can be captured accurately using an effective Hamiltonian constraint \cite{Taveras:2008ke} which captures underlying quantum dynamics very accurately \cite{Diener:2014mia}. With both holonomy and inverse scale factor corrections, the modified Friedmann equation is 
\begin{align}\label{effective}
 H^2  =\frac{8 \pi G}{3} \left[{\rho_{\Lambda}+ {\tilde{B}(v)}\rho_{\phi} } - \tilde{A}(v)\rho_{1}\right]\left[\frac{\tilde{A}(v)\rho_{2} - {\rho_{\Lambda}- {\tilde{B}(v)\rho_{\phi} }}}{\rho_{c}}\right],
\end{align}
with $\rho_{\Lambda} = \Lambda/(8\pi G)$ and $\rho_{\phi} = p_{\phi}^2/(2V^2)$ being the energy density of cosmological constant and massless scalar field, respectively, and $\rho_{c}  = {3}/({8 \pi G \gamma^2 \Delta })$ is critical energy density, and 
\begin{align}
\rho_{1} & =  - \rho_{c}  k  \chi,\\
\rho_{2} & = \rho_{c} \left(1- k \chi\right),\\
 \chi & = \sin^2 \bar \mu - (1+\gamma^2) \bar \mu^2,\\
 v & = K \left(\frac{6}{8 \pi \gamma l_{Pl}^2}\right)^{\frac{3}{2}} a^3.
\end{align}
with $ \Delta := 4 \sqrt{3} \pi \gamma l_{Pl}^2 = \bar \mu^2 a^2 $. Furthermore, $\tilde{A}(v)$ and $\tilde{B}(v)$ are the inverse scale factor corrections for gravitational and matter sector, respectively, given by \footnote{It should be pointed out that $\tilde{A}(v)$ and $\tilde{B}(v)$ terms have been redefined and they are not the original $A(v)$ and $B(v)$ terms in Ref. \cite{Ashtekar:2006es}.}
\begin{align}
    \tilde{A}(v) & = \frac{1}{2} \left|\left|v-1\right|- \left|v+1\right|\right|, \\
\tilde{B}(v) & = \left(\frac{3}{2}\right)^{3}   |v|^2 \left||v+1|^{1/3} - |v-1|^{1/3}\right|^3,
\end{align}
where $K = 2/3\sqrt{3\sqrt{3}}$ and $\gamma =  0.2375$ is Barbero-Immirzi parameter. One can check that Eq. (\ref{effective}) reduces into standard Friedmann equation including cosmological constant and massless scalar field at large volume limit where $\tilde{A}(v) \rightarrow 1$, $\tilde{B}(v)\rightarrow 1$ and $\Delta \rightarrow 0$. We see from the modified Friedmann equation that the trunarounds of scale factor can be obtained from $\rho_{\Lambda} + \tilde{B}(v)\rho_{\phi} = \tilde A(v) \rho_1$ and $\rho_{\Lambda} + \tilde{B}(v)\rho_{\phi} = \tilde A(v) \rho_2$. Since massless scalar field is proportional to $a^{-6}$, there may exist several distinct turnaround points depending on the value of cosmological constant and $p_{\phi}$ as we see in next section. However, the nature of turnaround, whether it is a bounce or a recollapse or in the Einstein static phase can be determined using Raychaudhuri equation. The Raychaudhuri equation including both holonomy and inverse scale factor corrections is given by  
\begin{align}\label{Raychaudhuri}
\nonumber \frac{\ddot a}{a}& = - \frac{4 \pi G}{3}\left(  \left({\tilde{A}(v)} - 3 v \tilde{A}^{\prime}(v)  \right) \left(\rho_{\Lambda}+ \rho_{\phi} \tilde{B}(v)\right) + 3 \tilde{A}(v) ( P_{\Lambda} + P_{\phi} ) \right) \\& \nonumber + \frac{16\pi G}{3} \left(\left(-\frac{1}{2} + \frac{3}{2} \tilde{A}(v)\right) \left(\rho_{\Lambda}+ \rho_{\phi} \tilde{B}(v)\right) + \frac{3}{2}(  P_{\Lambda} + P_{\phi})\right)\left(\frac{\rho_{\Lambda}+ \rho_{\phi} \tilde{B}(v)}{\rho_{c}} + {\tilde{A}(v)} k \chi\right)   \\&  \nonumber +  {\tilde{A}(v)} \frac{k\chi}{\gamma^2 \Delta}\left({\tilde{A}(v)} + 3 v \tilde{A}^{\prime}(v)   - \left(\tilde{A}(v)\right)^2\right)  \\ & \nonumber + \left(-1 + 2 {\tilde{A}(v)} - 3 \frac{v}{\tilde{A}(v)} \tilde{A}^{\prime}(v)\right) {\tilde{A}(v)} \frac{k\chi}{\gamma^2 \Delta} \left(\frac{\rho_{\Lambda}+ \rho_{\phi} \tilde{B}(v)}{\rho_{c}} + {\tilde{A}(v)} k \chi \right) \\ &   +  \left[2  \left( {\tilde{A}(v)} - \left({\tilde{A}(v)}\right)^2\right) \frac{k \chi}{\gamma^2 \Delta}-2 {\tilde{A}(v)} \frac{k \zeta}{\Delta \gamma^2}\right] \left( \frac{\rho_{\Lambda}+ \rho_{\phi} \tilde{B}(v) }{\rho_{c}} + {\tilde{A}(v)}k \chi - \frac{1}{2} {\tilde{A}(v)}\right),
\end{align}
with
\begin{align}
\zeta &= \sin^2 \bar \mu -  \bar \mu \sin\bar \mu \cos \bar\mu,
\end{align}
where $P_{\phi} = - \partial \mathcal{H}_{\phi}/\partial V $ with $\mathcal{H}_{\phi} = \tilde{B}(v)\rho_{\phi} V$. One can easily check that Eq. (\ref{Raychaudhuri}) reduces to Raychaudhuri equation in classical cosmology in large volume limit. However, in small scale factor regime, i.e., $v\ll 1$, $\tilde{A}(v) \sim v$ and $\tilde{B}(v) \sim v^{5}$, hence, all terms containing $\tilde{A}(v)$ and $\tilde{B}(v)$ terms in the RHS of Eq. (\ref{Raychaudhuri}) will be zero at zero scale factor. The only non-vanishing term is term with cosmological constant but at zero scale factor, we have $\ddot a = 0$. Hence, one can find that $\ddot a = \dot a = 0$ for $a = 0$ meaning that the universe is in Einstein static phase even if one considers cosmological constant with massless scalar field. We will see that this will have significant implication for tunneling wavefunction proposal in LQC.  

\section{effective minisuperspace potential}\label{III}

In one dimensional minisuperspace quantum cosmology, one can derive the effective minisupespace potential either from Hamiltonian constraint or Friedmann equation by an overall scaling by powers of scale factor \cite{Motaharfar:2022pjp}. However, one should point out that this analogy is true just for one dimensional minisuperspace quantum cosmology. Similarly, having the effective Friedmann equation (\ref{effective}), one can derive the effective minisuperspace potential including both holonomy and inverse scale factor corrections for cosmological constant with massless scalar field as follows
\begin{align}
    U_{eff}(a) &= -\frac{12}{8 \pi G} a^4  \left({{\rho_{\Lambda}+ {\tilde{B}(v)\rho_{\phi} }}}- \tilde{A}(v)\rho_{1}\right)\Big(\frac{\tilde{A}(v)\rho_{2}-{{\rho_{\Lambda} - {\tilde{B}(v)\rho_{\phi} }}}}{\rho_{c}}\Big),
\end{align}
while it reduces to the effective minisuperspace potential found in Wheeler-DeWitt quantum cosmology, Eq. (\ref{potential}), in large volume limit with $\rho = \rho_{\Lambda} + \rho_{\phi}$. In Ref. \cite{Motaharfar:2022pjp}, it was shown that in case of positive coamological constant, i.e., $\rho_{\phi}=0$, holonomy correction lifts the effective minisuperspace potential at zero scale factor excluding the possibility that the universe to be created out of nothing satisfying tunneling boundary conditions. Including the inverse scale factor correction, i.e., $A(v)$ term, it was found that the effective minisuperspace potential recovers its barrier shape, hence, the universe can tunnel from nothing into expanding universe or quantum cyclic universe satisfying tunneling boundary conditions. However, as we discussed in the introduction, the energy density of inflaton field becomes kinetically dominated at bounce whereby the effective minisuperspace potential will diverge at zero scale factor indicating that tunneling wavefunction is inconsistent with LQC. Moreover, those cyclic universes which are constructed by cosmological constant and matter content with $\omega \ge -1/3$ seems to be stable against the quantum decay to nothing which is in contradiction with results found in Ref. \cite{Mithani:2014jva}. To investigate these issues, we consider the universe filled with cosmological constant and massless scalar field with and without the inverse scale factor correction for massless scalar field.

\subsection{$\tilde{B}(v) = 1$}

\begin{figure}
    \centering
    \includegraphics[scale = 0.6]{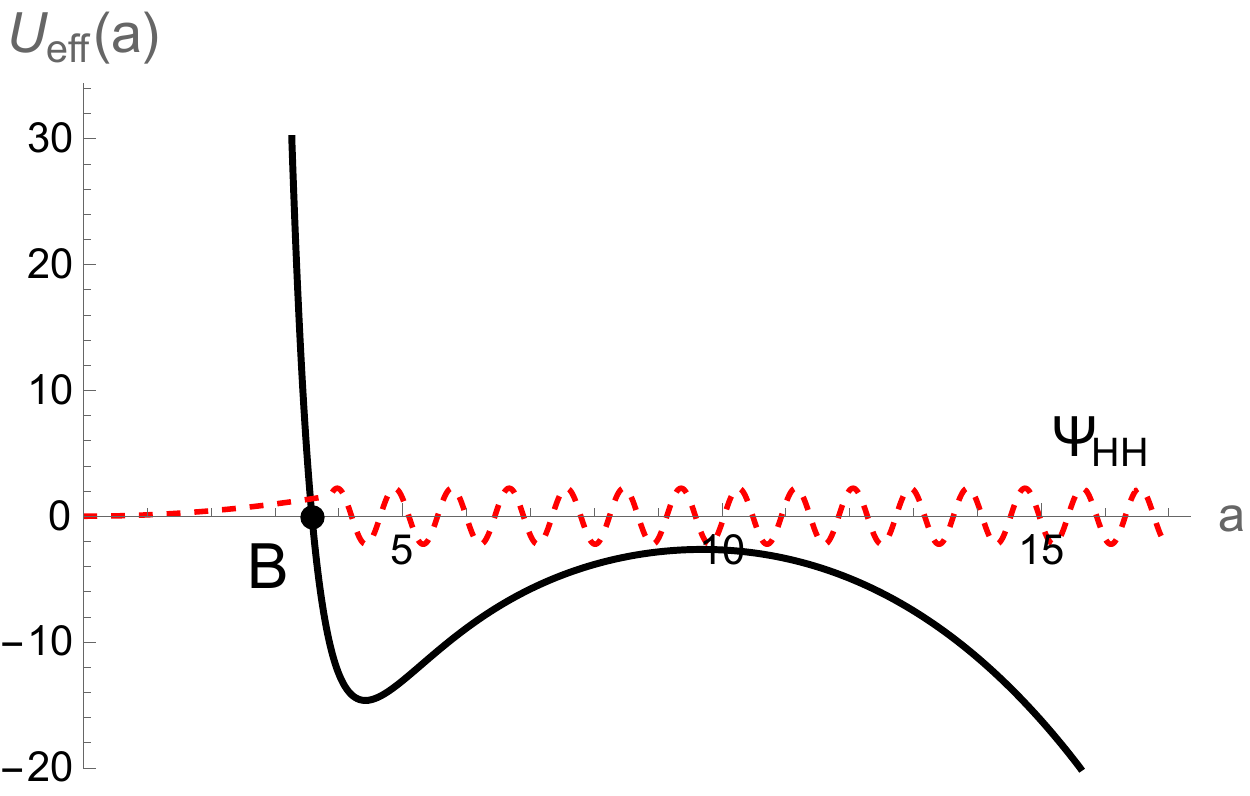}
    \includegraphics[scale = 0.6]{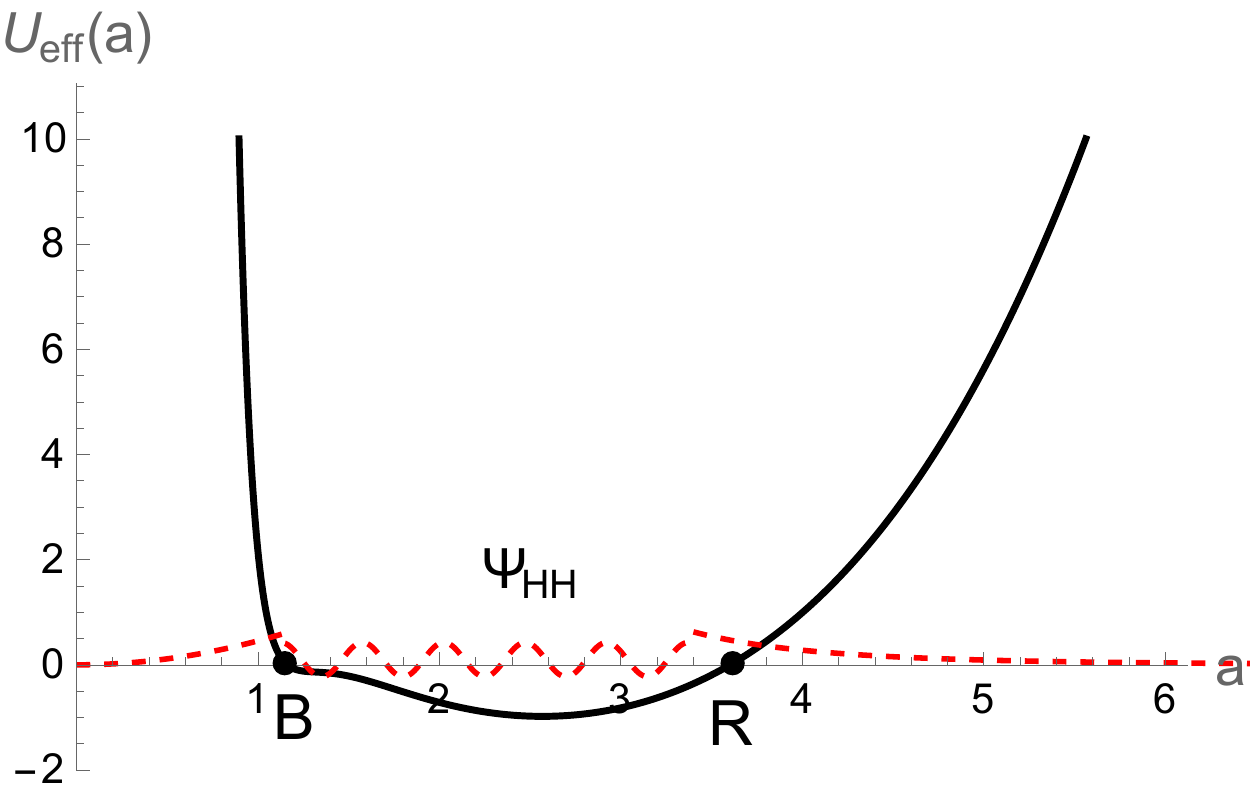}
    \caption{Effective minisuperspace potential including holonomy and inverse scale factor corrections (just $\tilde{A}(v)$ term) and schematic behavior of Hartle-Hawking wavefunction for $\Lambda = 0.03$ and $p_{\phi}= 600$  (left),  and $\Lambda = 11$ and $p_{\phi}= 35$ (right). Points B and R denotes the
bounce and recollapse turnaround points, respectively.}
    \label{fig2}
\end{figure}

\begin{figure}
    \centering
    \includegraphics[scale = 0.6]{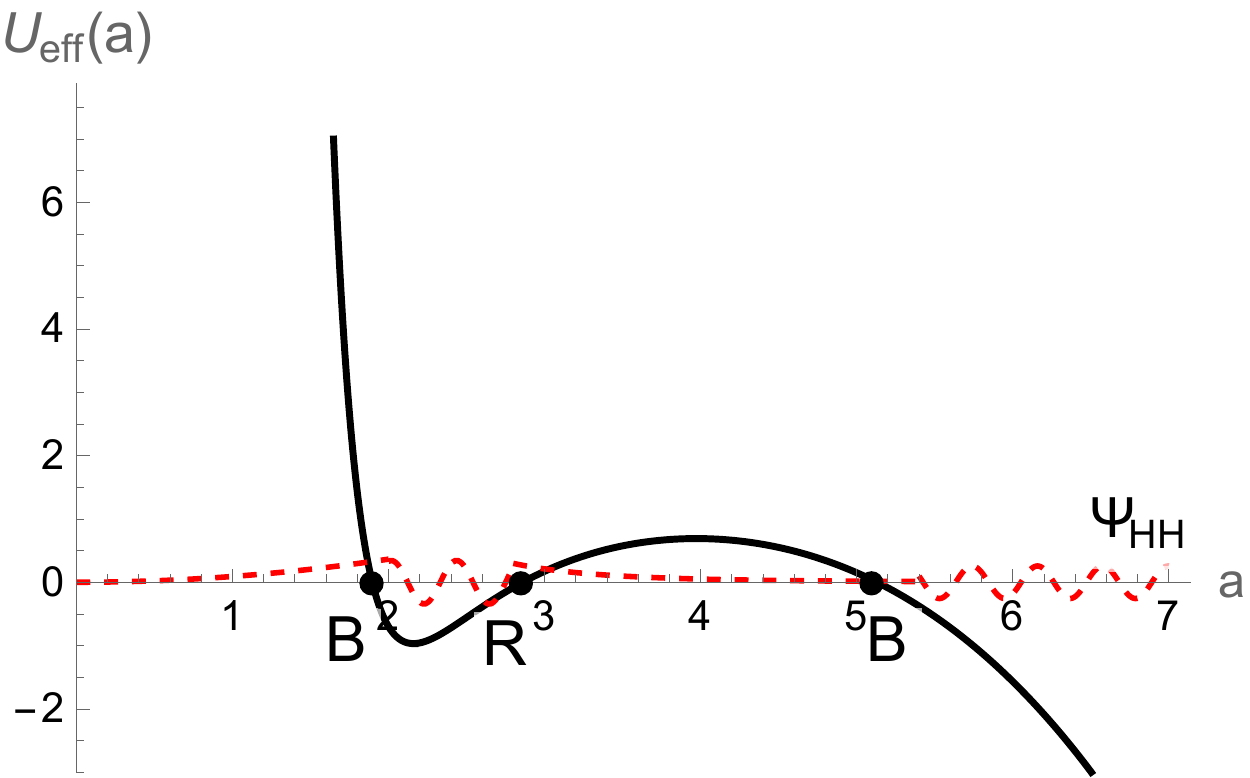}
   \includegraphics[scale = 0.6]{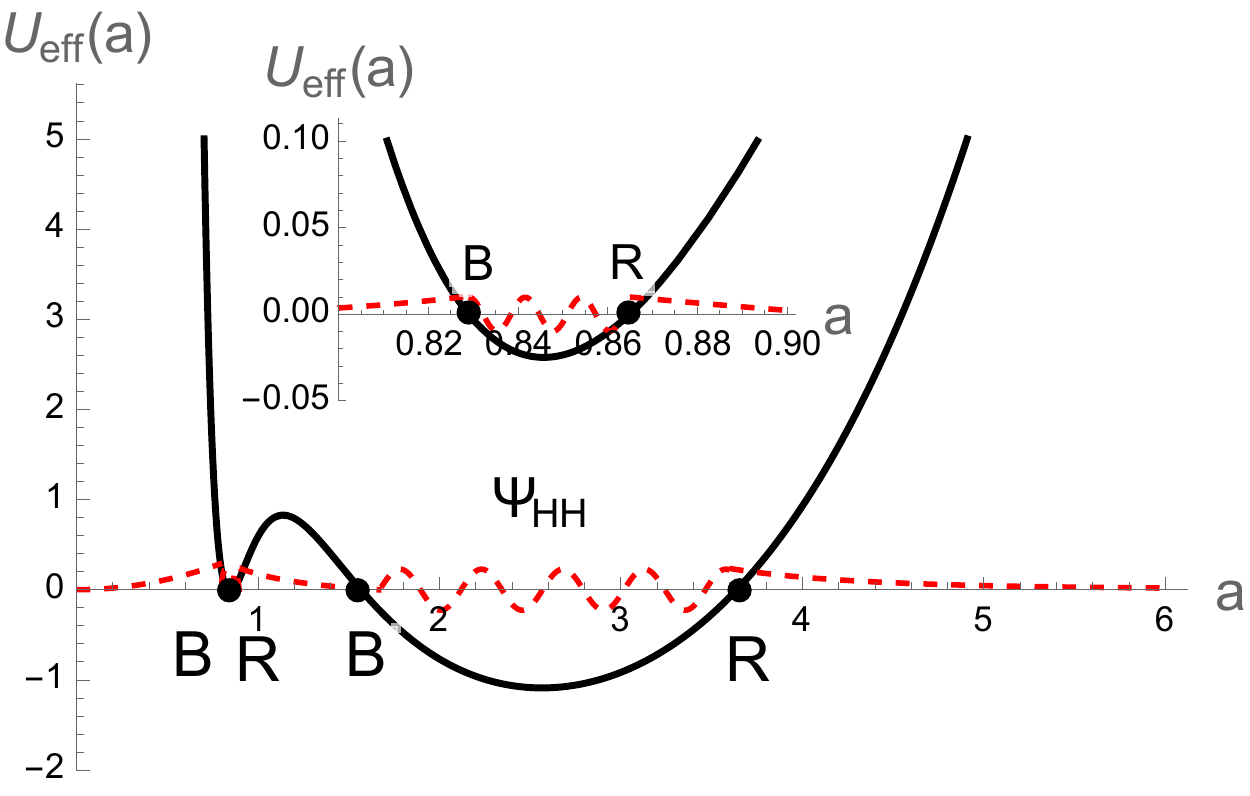}
    \caption{Effective minisuperspace potential including holonomy and inverse scale factor corrections (just $\tilde{A}(v)$ term) and schematic behavior of Hartle-Hawking wavefunction for $\Lambda = 0.2$ and $p_{\phi}= 110$  (left),  and $\Lambda = 11$ and $p_{\phi}= 20$ (right). Points B and R denotes the 
bounce and recollapse turnaround points, respectively.}
    \label{fig3}
\end{figure}

In this section, by ignoring the inverse scale factor correction for energy density of massless scalar field, i.e., $\tilde{B}(v) = 1$, we plotted the effective minisuperspace potential for four different cases in Fig. \ref{fig2} and \ref{fig3}. In fact, depending on the value of cosmological constant and $p_{\phi}$, the effective minisuperspace potential may have one, two, three or four turnaround points. However, the effective minisuperspace potential diverges at zero scale factor in all four cases meaning that the wavefunction should be decreasing towards the zero scale factor. Hence, the wavefunction cannot increase towards the zero scale factor, as a result of which the universe can be created from nothing using Hartle-Hawking boundary proposal (red dashed curve in Fig. \ref{fig2} and \ref{fig3}) rather Vilenkin proposal. In the left panel of Fig \ref{fig2}, we plotted the effective minisuperspace potential for $\Lambda = 0.03$ and $p_{\phi} = 600$ where it has just one bounce turnaround point, therefore, the universe is created out of nothing into classical expanding universe. While in the right panel of Fig \ref{fig2}, we used super-Planckian cosmological constant $\Lambda=11$ and $p_{\phi} =20$ due to which the universe recollapses at late time. However, this turnaround point has quantum nature, so the universe can tunnel from nothing into quantum cyclic universe in this case. In the left panel of Fig \ref{fig3}, we plotted the effective minsuperspace potential for $\Lambda = 0.1$ and $p_{\phi} = 110$ where it has three turnaround point. In fact, the universe can tunnel from nothing into cyclic universe which can play as seed for creating large expanding universe as it tunnels from the barrier (similar to what was found in Ref. \cite{Hertog:2021jyd}). Finally, in the right panel of Fig \ref{fig3}, we plotted the effective minisuperspace potential for $\Lambda=11$ and $p_{\phi} = 20$ where it has four turnaround points. In fact, the universe can tunnel from zero scale factor into the the first quantum cyclic universe while form there it can also tunnel to the second quantum cyclic universe as it recollapses. We conclude that if one ignores the effect of inverse scale factor correction for matter content, the cyclic universe is stable against quantum decay to vanishing size since the effective minisuperspace diverges at zero scale factor as we expected. 

\subsection{$\tilde{B}(v) \neq 1$}

\begin{figure}
    \centering
    \includegraphics[scale = 0.6]{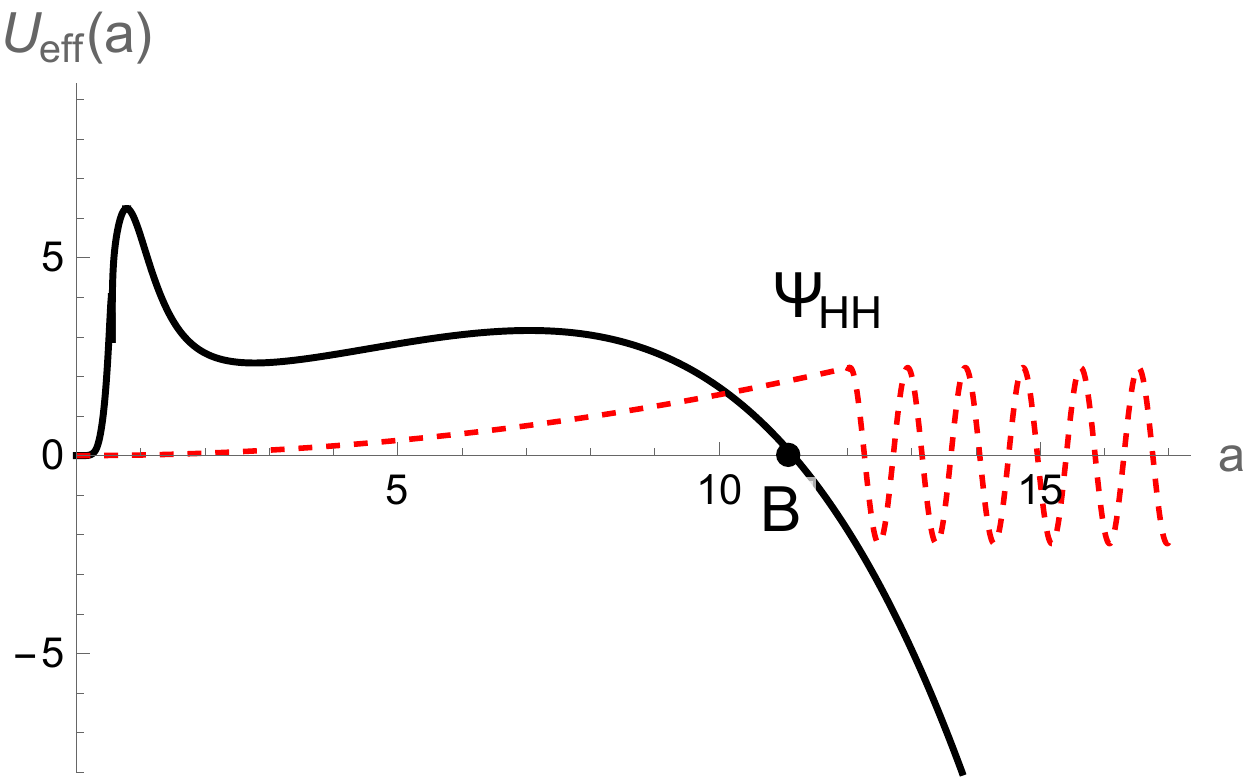}
    \includegraphics[scale = 0.6]{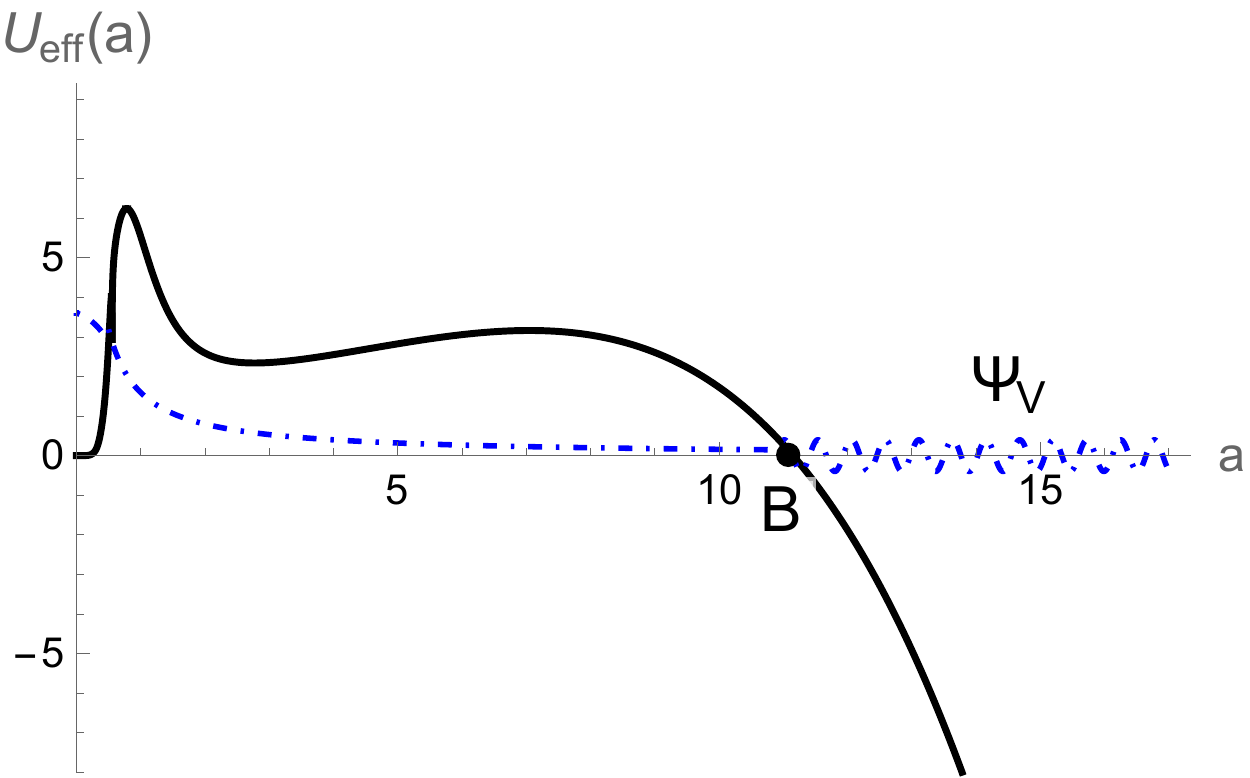}
    \caption{Effective minisuperspace potential including both holonomy and inverse scale factor corrections (both $\tilde{A}(v)$ and $\tilde{B}(v)$ terms) and schematic behavior of Hartle-Hawking wavefunction (left) and Vilenkin wave-function (right) for $\Lambda = 0.03$ and $p_{\phi} = 2$. Point B denotes the classical bounce turnaround point.}
    \label{fig4}
\end{figure}

\begin{figure}
    \centering
    \includegraphics[scale = 0.6]{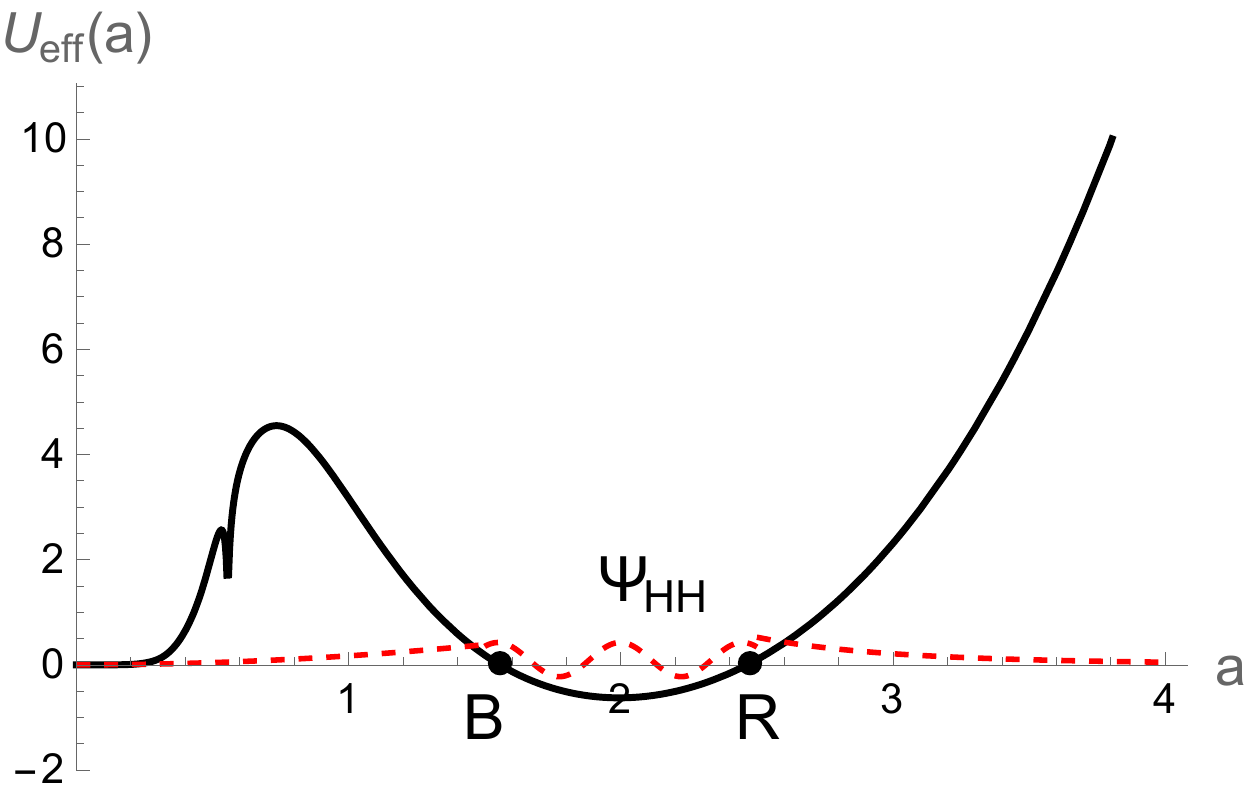}
    \includegraphics[scale = 0.6]{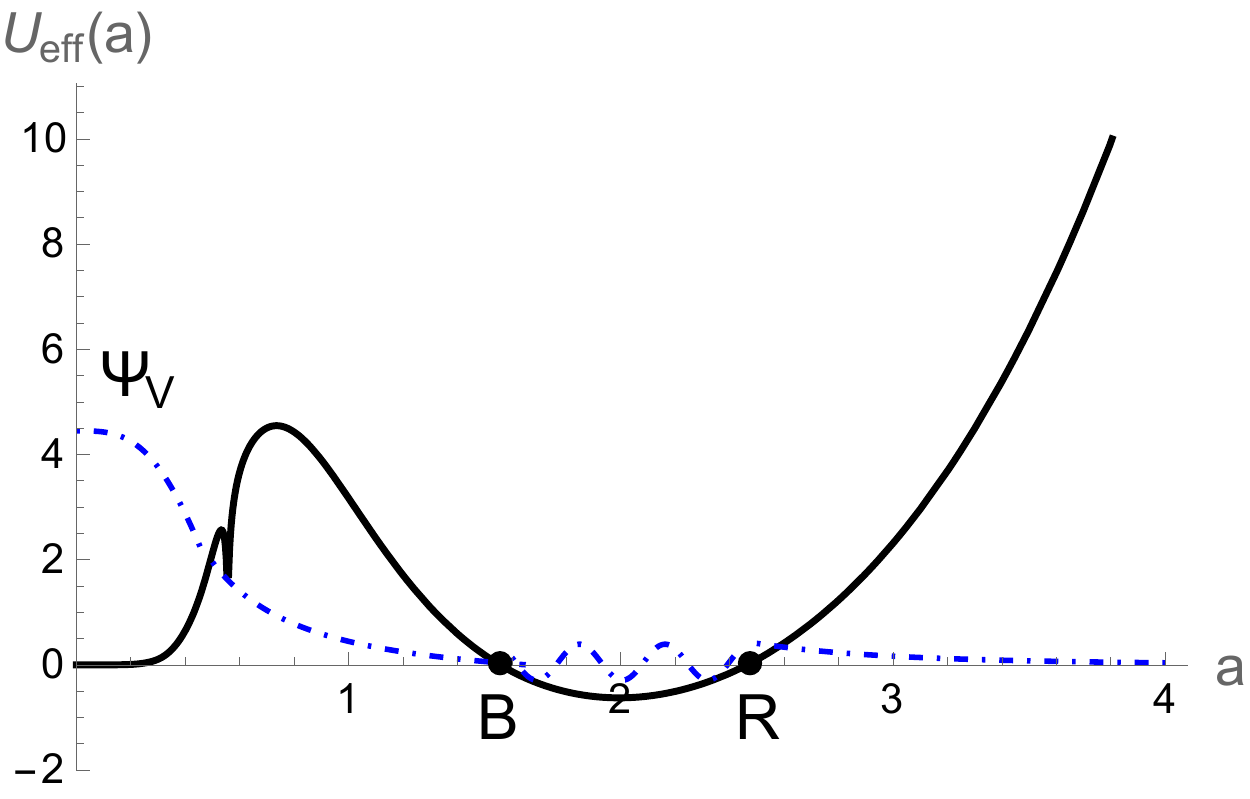}
    \caption{Effective minisuperspace potential including both holonomy and inverse scale factor corrections (both $\tilde{A}(v)$ and $\tilde{B}(v)$ terms) and schematic behavior of Hartle-Hawking wavefunction (left) and Vilenkin wave-function (right) for $\Lambda = 13$ and $p_{\phi} = 3$. Points B and R denotes the classical bounce and recollapse turnaround points, respectively.}
    \label{fig5}
\end{figure}

Although the spacetime is non-singular in the presence of holonomy and inverse scale factor corrections, one needs to include inverse scale factor correction for massless scalar field specially in studying tunneling wavefunction since the universe is deep in Plankian regime, i.e., $v\ll 1$. In this regime, $\tilde{B}(v) \sim v$ leading into no-trivial results. We plotted the effective minisuperspace potential in Fig. \ref{fig4} and Fig. \ref{fig4} for those value of cosmological constant for which we only have just one or two turnaround points. However, due to complicated behavior of $\tilde{B}(v)$ term, one can get several turnarounds points like what we discussed in previous section. Looking at Fig. \ref{fig4} and \ref{fig5}, one can see that the effective minisuperspace potential (black curve) recovers its barrier shape since $\tilde B(v) \rightarrow 0$ for $a = 0$ as it is the case in Wheeler-DeWitt quantum cosmology. Hence, the wavefunction can either pick up the decreasing or increasing wave mode under the barrier. Therefore, the universe can tunnel from nothing into classical expanding universe (Fig. \ref{fig4}) or quantum cylcic universe (Fig. \ref{fig5}) satisfying either tunneling or no-boundary boundary conditions. One should point out that going into zero scale factor is not in contradiction with the prediction of LQC that the spacetime is non-singular although it is in contradiction with the occurrence of bounce. On the other hand, since the universe is again able to tunnel back from bounce to zero scale factor, cyclic universe constructed from cosmological constant and perfect fluid with $\omega \ge -1/3$ is not stable against quantum decay to vanishing size which is consistent with the results found in Ref. \cite{Mithani:2014jva}.  

To reach this conclusion we used the effective Friedmann equation (\ref{effective}) which is accurate when the quantum fluctuations are small while quantum fluctuation are large near to big bang singularity for quantum tunneling. Therefore, the effective dynamics may not be trustable near to zero scale factor. However, the effects of such large quantum fluctuation on accuracy of effective equations were thoroughly investigated in Refs. \cite{Corichi:2011rt, Diener:2014hba, Ashtekar:2015iza} for flat spacetime in which they showed that such quantum fluctuation will change the upper bound on the maximum energy scale for which the bounce will occur in flat isotropic universe. In fact, it was found that for generalized Gaussian states, the maximum energy density should be replaced by 
\begin{align}
\tilde{\rho}_{c} = \rho_{c} \frac{e^{\frac{-\alpha^2}{\sigma^2}}\left(1+ \frac{\sigma^2}{4k_{0}^2}\right)}{\left(1+ \frac{\sigma^2 + \alpha^2}{4k_{0}^2}\right)}
\end{align}
where $\alpha = \sqrt{12\pi G}$ and $\tilde{\rho}_{c}<\rho_{c}$, and it approaches $\rho_{c}$ when $k_{0}\rightarrow \infty$ assuming dispersion relation $\sigma \sim \sqrt{2\alpha k_{0}}$ and $k_{0}\gg \alpha$. Since the massless scalar field become dominated at the bounce, therefore, one can neglect the effect of intrinsic curvature, so it is legitimate to assume the same effect for close LQC in the presence of large quantum fluctuations for the case studied here. Therefore, large quantum fluctuations only lower the energy scale for which the bounce occurs. In other words, such large quantum fluctuations will change the height of barrier whereby changing the rate of nucleation probability.

\section{Summary and Conclusion}

In this manuscript we investigated the possibility of creating the universe out of nothing via tunneling wavefunction proposal, considering loop quantum geometry effects while assuming the validity of effcteive dynamis in all regimes. Since inflaton energy density will be generally kinetically dominated at the bounce, we considered the cosmological constant together with massless scalar field to mimic the inflationary cosmology. We found that since the massless scalar field behaves as $a^{-6}$, it modifies the effective minisuperspace potential at small scale factor producing an infinite wall. Hence, the universe cannot tunnel from nothing into an expanding universe satisfying tunneling boundary conditions. However, in small scale factor regime one should consider the effect of inverse scale factor correction for matter component which are proportional to inverse of scale factor. We showed that including the inverse scale factor correction for massless scalar field, the effective minisuperspace potential recovers its barrier shape in small scale regime and tunneling wavefunction proposal can explain the initial condition for the universe. Although effective dynamics are valid only for small quantum fluctuations, using generalized Guassian states, we discussed that large quantum fluctuations only leads into lower energy scale for the bounce to happen, as a results of which the height of the barrier changes and the rate of nucleation probability also changes accordingly.

In addition, we also considered cyclic universe by choosing large cosmological constant due to which the universe recollapses at late time. It was shown in Ref. \cite{Motaharfar:2022pjp} that the universe can tunnel from nothing into cyclic universe for pure de Sitter universe including both holonomy and inverse scale factor corrections. However, as the universe recollpases it can also tunnel back to zero scale factor indicating quantum instability of cyclic universe in the context of LQC. In fact, this is true for any cyclic universe which is constructed using cosmological constant and matter component with $\omega<-1/3$. However, if one constructs cyclic universes with $\omega>-1/3$, such as massless scalar field, the effective minisuperspace potential diverges at zero scale factor whereby the universe cannot tunnel from zero scale factor to cyclic universe satisfying the tunneling boundary conditions. Accordingly, the universe cannot tunnel back from bounce to zero scale factor, therefore, the cyclic universe is stable against quantum decay to vanishing size. However, in small scale factor regime, one must also consider inverse scale factor correction for matter component. We showed that in this case the universe is again able to tunnel from zero scale factor into cyclic universe and tunnel back to zero scale factor as it recollapses again indicating quantum instability of cyclic universe in LQC. In fact, this is true for any cyclic universe which is constructed from cosmological constant with perfect fluid if we include the inverse scale factor corrections for matter content.

\begin{acknowledgements}
This work is supported by the NSF grant PHY-2110207.
\end{acknowledgements}

 \end{document}